\documentclass[12pt,preprint]{aastex}
\usepackage{natbib}

\def\OO#1{{\cal O}(#1)}

\shorttitle{eccentric harmonic aliasing of GJ 581d/g}
\shortauthors{G. Anglada-Escud\'e, R. I. Dawson}

\begin{document}

\title{Aliasing of the first eccentric harmonic : Is GJ 581g a genuine planet candidate?}

\author{Guillem Anglada-Escud\'e\altaffilmark{1}  \&
Rebekah I. Dawson\altaffilmark{2}
}

\email{guillem.anglada@gmail.com, rdawson@cfa.harvard.edu}

\altaffiltext{1}{Carnegie Institution of Washington, Department of Terrestrial Magnetism, 5241 Broad Branch Rd. NW, Washington D.C., 20015, USA}
\altaffiltext{2}{Harvard-Smithsonian Center for Astrophysics, 60 Garden St, MS-10, Cambridge, MA 02138}

\begin{abstract} 

The radial velocity (RV) method for detecting extrasolar planets has been the most
successful to date. The RV signal imprinted by a few Earth-mass planet around a cool
star is at the limit of the typical single measurement uncertainty obtained using
state-of-the-art spectrographs. This requires relying on statistics in order to
unearth signals buried below noise. Artifacts introduced by observing cadences can
produce spurious signals or mask genuine planets that should be easily detected
otherwise. Here we discuss a particularly confusing statistical degeneracy resulting
from the yearly aliasing of the first eccentric harmonic of an already-detected
planet. This problem came sharply into focus after the recent announcement of the
detection of a 3.1 Earth mass planet candidate in the habitable zone of the nearby
low mass star GJ 581. The orbital period of the new candidate planet (GJ 581$g$)
corresponds to an alias of the first eccentric harmonic of a previously reported
planet, GJ 581$d$. Although the star is stable, the combination of the observing
cadence and the presence of multiple planets can cause period misinterpretations. In
this work, we determine whether the detection of GJ 581$g$ is justified given this
degeneracy. We also discuss the implications of our analysis for the recent Bayesian
studies of the same dataset, which failed to confirm the existence of the new planet.
Performing a number of statistical tests, we show that, despite some caveats, the
existence of GJ 581$g$ remains the most likely orbital solution to the currently
available RV data.

\end{abstract}

\keywords{
Stars: individual (GJ 581) --- 
planetary systems ---
Techniques: radial velocities ---
Methods: data analysis} 

\section{Introduction} 

The recently reported planet candidate around the nearby M dwarf GJ 581
\citep[][hereafter V10]{vogt:2010} has generated much public enthusiasm and a similar
amount of skepticism within part of the scientific community \citep{iau:torino}. If
confirmed, it will be the first planet potentially capable of hosting life as we know
it now \citep{kasting:1993,braun:2011,heng:2010}. The possible existence of this
planet was announced based on analysis of the new HIRES/Keck precision RV
measurements combined with HARPS/ESO data published by \citet{mayor:2009} (hereafter
M09). V10 reported that the candidate planet GJ 581$g$ (hereafter \textit{planet g})
has a minimum mass of $3.1$ $M_\oplus$ and a period of 36.5 days. The data for GJ 581
contain the signal of at least 4 other low mass planets (planets $b-e$). Dynamical
studies showed that their masses cannot be larger than $\sim$1.4 times the reported
minimum masses (M09 and V10).

Of particular interest to this work is GJ 581d. The planet has a period of 67 days and
a previously reported eccentricity of $\sim 0.4$ (M09). Using only the HARPS data
(M09), \citet{anglada:2010a} noted that the eccentric orbit of GJ 581$d$ could
disguise the signal of a low mass companion at half its period, similar to the
candidate planet $g$ now reported by V10. Both eccentric planets and pairs in or near
2:1 mean-motion resonance have been discovered by RV and transit surveys\footnote{see
exoplanet.eu for an up-to-date census}. Each case implies a very different dynamical
history so the distinction is important in understanding evolution of planetary
systems.

The available RV data of GJ 581 is strongly affected by yearly aliasing
\citep[][hereafter DF10]{dawson:2010}. In Section \ref{sec:aliases}, we explain this
aliasing and demonstrate how the eccentricity of a planet can be confused with an
additional planet at half its period. In Section \ref{sec:degeneracy}, we show that
planet GJ 581$g$ has an orbital frequency that aliases to the eccentric harmonic of
planet $d$ and therefore could be an artifact of the sampling cadence. In Section
\ref{sec:analysis}, we reevaluate the significance of planet $g$ in the presence of
this degeneracy by allowing all the planetary orbits to be eccentric. In the same
section, we quantify the probability of confusing planet $g$ with the eccentric
harmonic of planet $d$. In our analysis, we encountered some caveats about the
reality and uniqueness of the reported signal. In Section \ref{sec:caveats}, we
describe these caveats and their relation with the failure of recent Bayesian studies
to confirm the presence of planet $g$ \citep{tuomi:2011,gregory:2011}. Given that the
6-planet solution proposed by V10 is the most significant one that is also
physically allowed, we conclude that the presence of planet $g$ remains supported by
the data.

\section{Sampling cadence and aliases}\label{sec:aliases}

\begin{figure}[tb] 
   \centering
   \includegraphics[width=120mm,clip]{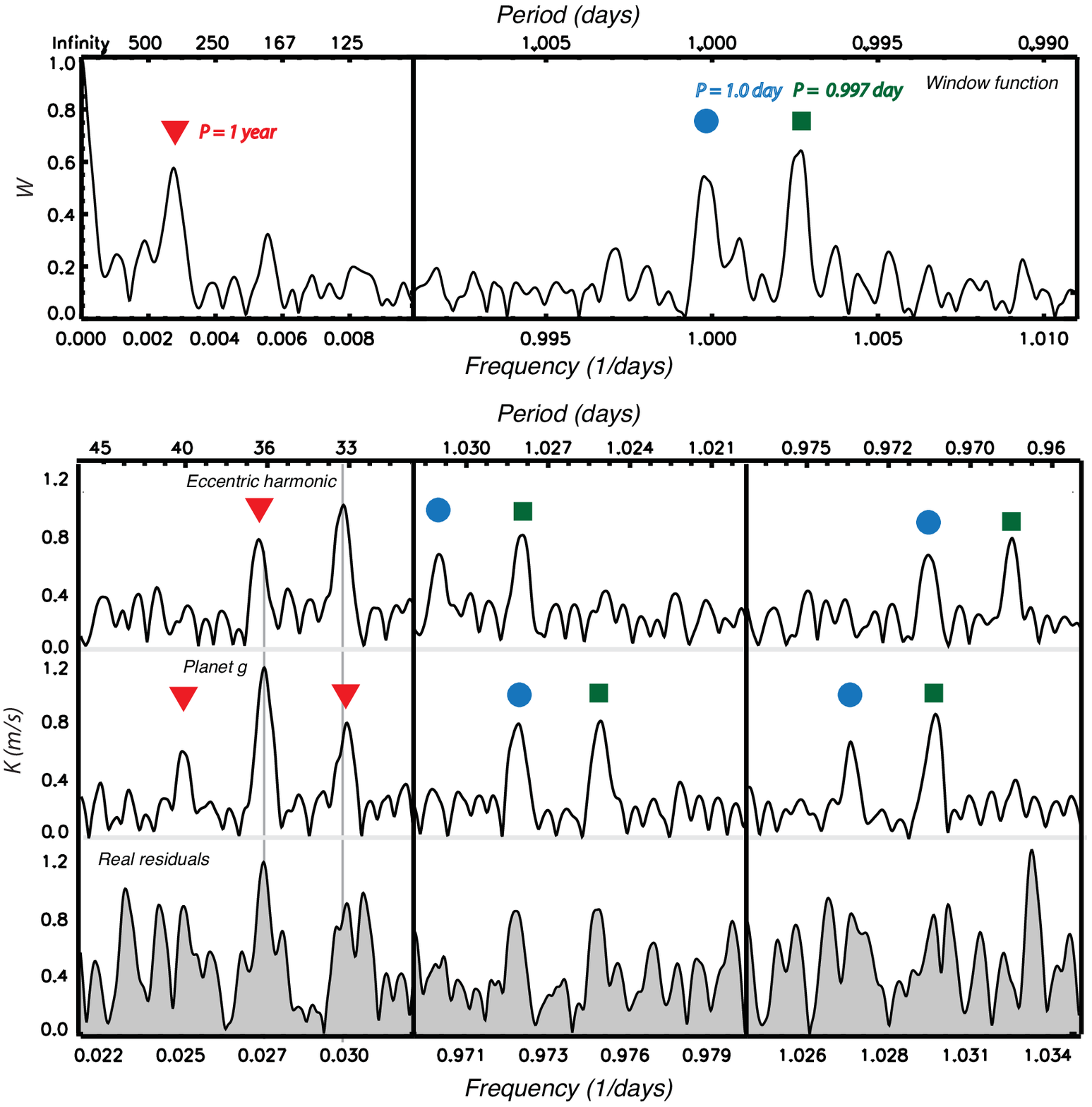}
   
\caption{
Upper panel: Modulus of the spectral window function including HARPS (M09) and HIRES
(V10) data. The top right panel shows both the sidereal and solar day as
significant sampling frequencies. The top left panel shows that the 1
year$^{-1}$ frequency will also create aliases. Bottom panel: Power spectra of
the residuals to a 4-planet fit with eccentricities preliminarily fixed at 0
(bottom), a noiseless synthetic planet $g$ (center), and a noiseless synthetic
signal at the frequency of the eccentric harmonic (top). The aliases are marked
with symbols corresponding to the upper panel. 
}


\label{fig:window}

\end{figure}

The sampling cadence of a periodic signal can alter its apparent period. An
\textit{alias} is a spurious periodicity generated by finite sampling of a real
signal. As an extreme example, if a function of period $P$ is measured at regular
intervals $\Delta t=P$, it will appear constant to the observer. In the case of even
sampling, the maximum frequency that can be probed without ambiguity is half the
sampling frequency (the Nyquist frequency). However, this frequency cannot be
unambiguously determined in the case of uneven sampling \citep{eyer:1999}. Instead,
one can identify characteristic sampling frequencies by computing the 
\textit{spectral window function} $W$, which depends only on the observation instants.
The modulus of $W$ ranges from 0 to 1 and peaks at the sampling frequencies that cause
the most severe aliases. For a real signal of frequency $f = 1/P$ and a characteristic
sampling frequency $f_s$, aliases appear at $\left|f\pm f_s\right|$. Figure
\ref{fig:window} shows the window function including all the HARPS+HIRES data
available for GJ 581 (M09 and V10). Optical astronomical measurements occur only at
night and only during the season when the star is observable. The night time cadence
introduces two sampling frequencies close to 1 day$^{-1}$ (the solar and sidereal day,
top right panel), and the seasonal availability generates a 1 year peak (top left
panel).  When the signal--to--noise is low, coherent addition of noise can produce
higher power at the aliases than at the real period (see DF10). Since the announcement
of the first planet candidates around GJ 581 by \citet{bonfils:2005}, the yearly alias
has significantly affected the determination of the orbital periods of these planets. 
For example, the first period proposed for the super-Earth GJ 581d was 84 days 
\citep{udry:2007}, a yearly alias of the currently accepted period of 67 days. 

\begin{figure}[tb] 
   \centering
   \includegraphics[width=130mm,clip]{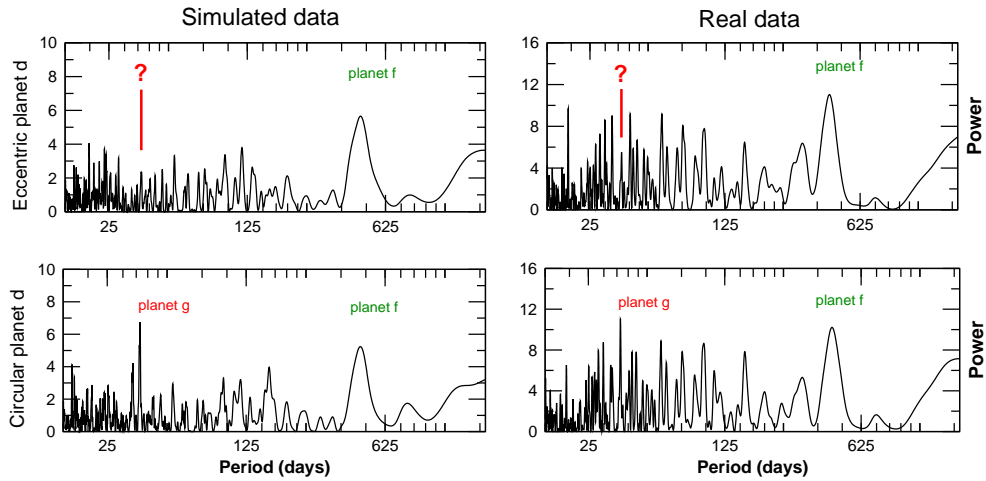}
\caption{ 
Periodograms of the residuals to a 4-planet fit containing the planets
in M09. Top: periodograms assuming an eccentric planet $d$.  Bottom:
periodograms of the same datasets, fixing $e_d$ at 0. The left
panels correspond to synthetic data containg the six planets announced by V10.
The right panels are periodograms of the residuals of actual data. The 36.5 day
signal will be missed if the eccentricity of planet $d$ is prematurely adjusted
(red question marks).
}\label{fig:aliasing}
\end{figure}

The other ingredient to the degeneracy is that the RV signal due to a planet's
Keplerian motion can be expressed as a series in powers of its orbital eccentricity,
$e$. To first order in eccentricity, a Keplerian RV signal of amplitude $K$ and period
$P$ can be written as

\begin{eqnarray}
v_{r}(t) = \gamma_{\rm I} + 
K \cos \left(\frac{2\pi t}{P} + \phi_0\right) + 
Ke \cos \left(\frac{2\pi t}{P/2} + \phi_1\right) + 
\OO {Ke^2}\,,
\label{eq:RV}
\end{eqnarray}

\noindent where $\phi_0$ and $\phi_1$ are functions of the initial mean anomaly and
the argument of the periastron, and $\gamma_{\rm I}$ is a constant offset that can
be different for each instrument. Eqn. \ref{eq:RV} implies that in a system with
a planet of period $P$, an additional body with period $P/2$ is statistically
indistinguishable unless the next term in the expansion can be measured. We call the
term proportional to $Ke$ the ``first eccentric harmonic.'' Due to statistical
uncertainties, finite sampling, and aliasing, two planets can be misinterpreted as a
single eccentric planet even when their period ratio is not exactly 2.

\section{Degeneracy between planets GJ 581d and g}\label{sec:degeneracy}


As shown by M09, the period of planet $d$ is $\sim 67$ days. Therefore, the first
eccentric harmonic of this planet has a period of $\sim 33.5$ days ($f = 0.02985$
day$^{-1}$).  The period of the newly found planet $g$ is $36.5$ days, and its
aliases should appear at $33.18$ and $40.55$ days ($|f = 0.02985$ day$^{-1} \pm
1/$yr$|$). Because a yearly alias of planet $g$ falls very close to the eccentric
harmonic of planet $d$, planet $g$'s signal will be partially absorbed by the
eccentricity of planet $d$. To illustrate this, we carry out the following experiment.
We generate a synthetic dataset, the exact 6-planet solution from V10 (all circular
orbits) sampled at the times of the HARPS and HIRES datasets and with Gaussian noise
added  (1.9 and 2.7 m/s respectively). Then, we compute the periodogram of the
residuals of a 4-planet fit to this synthetic dataset (Fig. \ref{fig:aliasing}).
Even though planet $g$ is included in the simulated data, its peak will disappear
if planet $d$ is allowed to be eccentric (top left). On the other hand, planet $g$
will appear prominently if the eccentricity of planet $d$ is preliminarily fixed at 0
(bottom left). The same behavior is observed for the real data (right panels). As a
general rule in modeling RV data sets, it is safer to search for significant
periodicities first (i.e. preliminarily fixing the eccentricities at 0) and defer
determining which model is preferred to a later stage (i.e. an eccentric
planet vs. a two planet model).


In DF10, a test was developed to qualitatively assess which period is more likely in
the presence of signal aliasing. The test consists of the following steps: (1)
generate noiseless synthetic datasets of the signals under study, (2) compute their
power spectra at the regions where the most prominent features occur (the nominal
period and strongest aliases), and (3) compare them with the power spectrum of the
real data (i.e. for GJ 581, the residuals to the 4-planet fit with eccentricities
preliminarily fixed at 0).  By \textit{power spectrum} we mean the amplitude of the
sinusoidal function that best fits the data at each frequency (see DF10 for further
details). 


We generate a synthetic dataset for each of the two candidate periods (33.5 days and
36.5 days). The bottom panels of Figure \ref{fig:window} show the power spectra of
the real and the synthetic datasets at the nominal periods and their most prominent
aliases. According to \citet{lomb:1976}, ``If there is a satisfactory match between
an observed spectrum and a noise-free spectrum of period $P$, then $P$ is the true
period". While the eccentric harmonic (33.5 days) and its aliases fail to mimic the
features observed in the real data (especially the daily aliases in the central
panels), the candidate signal at 36.5 days does a fairly good job of reproducing most
of them. Random noise modifies the power balance between peaks, so one should
understand this as a first qualitative assessment. The probability of confusing
planet $g$ with the eccentric harmonic of planet $d$ is quantified in Section
\ref{sec:confusion}.

\section{Detailed analysis}\label{sec:analysis}

\subsection{Preliminary period search}  

In order to indentify promising low-amplitide periodicities, first we
sequentially subtract the four most significant signals that coincide with those
reported by M09 (using the systemic interface \citep{systemic}). As discussed in
Section \ref{sec:degeneracy}, we preliminarily fix the eccentricities at 0. All
parameters are refined each time a planet is added. The periodogram
\citep{anglada:2010b} of the residuals to the 4-planet model is then computed.
Both the 36.5 and 433 day signals appear prominently, see bottom right panel of
Fig.\ref{fig:aliasing}. A \textit{detection} False Alarm Probability (FAP) is
obtained through a Monte Carlo approach using 10$^4$ synthetic realizations of
the data. Each synthetic dataset is created by scrambling the residuals to the
four--planet fit, keeping the same observing epochs and instrument membership.
The detection FAP is the fraction of times we obtain a signal with a higher
power than the one in the real residuals. We find that both \textit{detection}
FAP are below 0.5\%. After subtracting these two signals (36.5 and 433 days), no
additional peaks are found with a detection FAP under 1\%. These FAP are for
detection of sinusoid--like signals only. The actual significances for the
proposed new candidates are computed in the next section.

\subsection{Statistical significance of GJ 581g} \label{sec:significance}

Recently, \citet{gregory:2011} and \citet {tuomi:2011} indicated that planet $g$ is
not robustly confirmed when Bayesian analysis is applied. To address this issue, we
consider in this section the general case where the orbits of the already-detected
planets are allowed to be eccentric and compute the definitive FAP analytically using
a classical Frequentist approach. Because the uncertainties in RV measurements are
difficult to quantify, the $\chi^2$ statistic cannot be used to evaluate the
goodness-of-fit in an absolute sense \citep{andrae:2011}. However, the $\chi^2$ can
still be used in a differential way, i.e., to determine whether or not the addition of
a planet is justified given the improvement of the $\chi^2= \sum_i^{\rm
obs}r_i^2/\epsilon^2_i$ ($r_i$ are the residuals with respect to a proposed model and
$\epsilon_i$ are the uncertainties). In this respect, we apply a version of the Fisher
F-ratio test proposed by \citet{cumming:2008} that allows the addition of any number
of free parameters to the null hypothesis and to the model being tested. This requires
the computation of the F-ratio as

\begin{eqnarray} 
F =\frac{(\chi^2_{\rm null}-\chi^2_{+})/(k_{+}-k_{\rm null})}{
\chi^2_{+}/(N_{\rm obs}-k_{+})} \label{eq:fratio} \end{eqnarray} 

\noindent  where $k_{\rm null}$ and $k_+$ are the number of free parameters of the
null hypothesis and the model to be tested respectively. The F--ratio follows an
F--probability distribution with $(k_{+}-k_{\rm null})$ and $(N_{\rm obs}-k_{+})$
degrees of freedom \citep{fdist:1995}. The null hypothesis includes the M09 planets
(planets $b, c, d, e$) and planet $f$, all with fully Keplerian orbits. The
$\chi^2_{+}$ is obtained by adding a new planet on a $\sim$36.5 day circular orbit
(requiring two extra parameters) and readjusting all the free parameters. Using the
values in the two last columns of Table \ref{tab:stats}, the obtained probability of
improvement-by-chance at that particular frequency is $p = 4.41\times 10^{-5}$.  Now
one has to consider the number $M$ of independent frequencies that could also
generate a spurious improvement.  According to V10, M is $2525$ for the HARPS+HIRES
combined dataset. Therefore, the definitive FAP for planet $g$ becomes
FAP$=1-(1-p)^M$\citep{cumming:2008} and amounts to $0.11\%$. For completeness, we add
the FAP of planet $f$ ($\sim 0.03\%$) to Table \ref{tab:stats}, using the Keplerian
4-planet solution (M09) as the null hypothesis.

The F-ratio test can also be used to evaluate the significance of the eccentricity
compared to a circular orbit (the null hypothesis). We find that the eccentricities
of the larger amplitude planets (i.e. $b, c, d$ and $e$) are well-constrained but
compatible with $0$. For the low amplitude candidates ($f$ and $g$), the
eccentricities are unconstrained. Table \ref{tab:solution} provides the full orbital
solution.


\subsection{Probability of confusion}\label{sec:confusion}

At this point, we have assessed the FAP of the planet detection. Now we need to
quantify the probability that the eccentric harmonic of planet $d$ could inflate the
signal at 36.5 days by an unfortunate combination of random errors. We generate $10^4$
synthetic datasets by scrambling the residuals to the 6-planet fit and injecting a
signal at 33.5 days (amplitude $1.29$ m/s). The periodogram of each dataset is then
computed. We define the \textit{probability of confusion} as the number of times the
period of the highest peak does not lie within 1 day of the injected signal, divided
by the total number of trials. For the $33.5$ days, we find this probability is
$0.34\%$. In $0.12$\% of trials, the highest peak is found near the $36.5$ days alias.
Performing the same test by injecting the 36.5 days signal, we obtain a $0.5\%$
probability of confusion (with $0.2\%$ of trials corresponding to the 33.5 days
alias). As a final check, we repeat the same experiment using random noise instead of
scrambling the residuals ($\sigma = $ 1.8 m/s for HARPS, and $\sigma = $ 3.0 m/s for
HIRES). Similarly, low confusion rates ($\sim$ 0.5\%) are obtained. In every case, the
probabilities are below $1\%$, supporting the conclusion of the previous section:
\emph{the probability of confusing planet $g$ with the eccentric harmonic of planet $d$
is very low}. As a final check, we apply the same test using only the HARPS residuals.
In more than $50\%$ of the trials, the highest peak is none of the injected signals,
confirming that the data in M09 was not sensitive enough to distinguish between cases.

\begin{deluxetable}{lccccc}  
\tablecolumns{6}
\tablecaption{Statistical quantities required to evaluate
the F-ratio in Eqn. \ref{eq:fratio}}
\tablehead{   
  \colhead{Parameter} &
  \colhead{Four planets} &
  \colhead{Four planets} &
  \colhead{ } &
  \colhead{Five planets} &
  \colhead{Five planets} 
  \\
  \colhead{} &
  \colhead{M09 solution} &
  \colhead{+ circular f} &
  \colhead{ } &
  \colhead{ } &
  \colhead{+ circular g}
}
\startdata
Planets included\tablenotemark{a}  &  bced            & bcde(f)             &  & bcdef	              &  bcdef(g)              \\
Eccentric orbits  &  4                   & 4                   &  & 5		      &  5		                       \\
Circular orbits   &  0                   & 1                   &  & 0		      &  1		                       \\
Eccentricity \\ of planet $d$     &  $0.5$               & $0.4$               &  & $0.4$                &  $\sim 0.1$           \\
\hline 
                                                                                             			     	       \\
RMS (m/s)                      &    2.32  	      &  2.21		    &  &  2.20		      &	 2.10		       \\
Free parameters                &    22  	      &  24		    &  &  27		      &	 29		       \\
$N_{\rm obs}$                  &    241 	      &  241		    &  &  241		      &	 241		       \\
$\chi^2$                       &    707.7             &  611.8              &  &  602.5               &	 524.8  	       \\
                                                                \hline                              			       \\
F ratio                        &                      &   17.0              &  &  		      &	 15.69  	       \\ 
FAP                            &                      &   0.03\%            &  &  		      &	  0.11\%	       \\
\enddata
\label{tab:stats}

\tablenotetext{a}{A planet name in parentheses indicates that this solution was
used to compute the FAP of that planet. The number of free
parameters is obtained as follows: 2 RV offsets (one per instrument), 5 for each
Keplerian planet, and 2 for the circular orbit to be
tested.}

\end{deluxetable}



\subsection{Caveats}\label{sec:caveats}

During our analysis, we encountered a number of caveats that will require
further investigation. Recent analyses by other authors using Bayesian methods
\citep[i.e.][]{gregory:2011,tuomi:2011} seem to contradict the conclusions
reached here. Thus we discuss some of these caveats and possible lines of
inquiry.

\textbf{Significance of spurious periodicities.} We found a number of alternative
periodicities for planets $f$ and $g$ that yield alternative 6-planet fits with
similar significance to the solution announced by V10. Potential periods for planets
$f$ and $g$ include 25.0, 26.8, 30.6, 47.8, 54.6, 59.3, 71.4, 76.4 and 97.9 days.
These alternative solutions were found by computing a two-dimensional periodogram
(i.e. two periodic signals are adjusted simultaneously on a grid) of the residuals of
the 4-planet circular solution and refining all Keplerian parameters around the areas
of lowest $\chi^2$ \citep{mark:2009}. However, a case-by-case investigation indicated
that all correspond to orbital configurations with high eccentricities that suffer
from several orbital crossings, making them dynamically unstable on very short
timescales.  To ultimately decide if such solutions were acceptable, we fixed the
involved eccentricities to lower values, adjusting all other parameters in the
process. The resulting orbital fits were poorer than the 6-planet solution from V10
(even assuming circular orbits for planets $f$ and $g$).  The significance of these
unphysical solutions raises the caveat that planet $g$ could be a physically-possible
but spurious signal. As discussed by \citet{tuomi:2011}, Bayesian analysis methods do
not yet apply down-weighting of unphysical configurations. Because such orbits have
high significance, they will be oversampled, downgrading the likelihood of physically
allowed orbits. This may explain the apparent contradiction of the recent Bayesian
analysis with our conclusions.

\textbf{Hidden planets}. While planets $b$ and $c$ are detectable in both datasets,
planets $d$ and $e$ are not obvious in the HIRES data alone (as also noted by V10).
This indicates that some signals may be consistent with a given dataset, yet not
independently detectable due to sampling issues. As a similar case, GJ 876$e$ was a
strong signal in the HIRES dataset yet undetectable with HARPS \citep{rivera:2010}.
Because the completeness of on-going RV surveys can be affected as a result, this
caveat requires further investigation.

\textbf{Noise unknowns}. We have chosen statistical tools that are as insensitive as
possible to assumptions about noise: the F-ratio test and Monte Carlo scrambling of
the residuals. Still, we recognize that it is not fully understood how systematic
effects (e.g. stellar jitter) impact the sensitivity to low
amplitude signals. A Bayesian approach with a \textit{noise parameter} has been
proposed to solve this problem \citep{tuomi:2011}. However, we think that further
tests are required to assess the sensitivity of Bayesian methods in the presence of
unknown systematic noise and low amplitude signals.

\section{Conclusions}

The eccentric harmonic of planet $d$ coincides with a yearly alias of the newly reported
planet $g$, meaning that both signals are correlated and that a premature Keplerian
fit to planet $d$ prevents the detection of planet $g$. We have found a number of
unphysical solutions that fit the data just as well. Still, the proposed pair of
planets $f$ and $g$ remains the only physically viable solution that significantly
improves the 4-planet fit. Thus we are compelled to conclude that the presence of GJ
581$g$ is well supported by the available data. The ultimate confirmation will
require additional RV measurements and reanalysis of the data in a very convincing
way. To mitigate yearly aliasing, we encourage observations at more extreme parallax
factors.

We end with two cautionary notes. First, whether or not planet $g$ exists, the same
cadence issues may be present in other datasets. We have found that significant
periodicities in one dataset do not appear in another and that several unphysical
models can appear significant. If future observations rule out the existence of
planet $g$, the fact that it passes the standard, widely-used statistical tests would
bode ill for other low-amplitude planet candidates. We remark that statistical
significance tests can be very sensitive to assumptions by the authors
(including this work). Bayesian methods may provide stricter confidence
level estimates but additional testing is required to ensure that they are not
over-conservative in the low signal--to--noise regime \citep{jenkins:2011}. Bayesian
methods may also need to demonstrate how their conclusions are impacted by sampling
issues and the inclusion of dynamical constrains in the likelihood sampling process.

Second, the eccentricity of a long period giant planet may cause spurious low
amplitude signals mimicking habitable planets around Sun-like stars. Consider
Jupiter as an example: its eccentric harmonic would have a period of 5.9
years (or f = 0.00046162 days$^{-1}$). The corresponding yearly aliases
would be at $|f\pm$0.0027378 days $^{-1}|$, giving candidate signals at 439 and
313 days. Complementarily, a genuine planet candidate can be missed if the
eccentricity of an outer planet is adjusted prematurely. Aliasing tests on the
eccentric harmonics of detected planets might be necessary in future claims of
very low-amplitude signals.

\textbf{Acknowledgements} This work was funded by the Carnegie Fellowship
Postdoctoral Program and the National Science Foundation Graduate Research
Fellowship. We thank P. Butler, and S. Vogt for early access to the RV data. We
thank A. Boss, J. Chanam\'e,  J. Dunlap, D. Fabrycky, N. Haghighipour, M. Lopez-Morales, and N. Moskovitz
for useful comments and discussions.

\bibliographystyle{apj}

\begin{deluxetable}{lrrrr}  
\tablecolumns{5}
\tablecaption{Keplerian solution including 6 planets (V10).
Parameter values correspond to the best $\chi^2$ solution.
Uncertainties have been obtained using the Markov Chain Monte Carlo
algorithm included in the \textit{systemic} interface \citep{systemic}.
}
\tablehead{  
\colhead{Parameter} & \colhead{b}& \colhead{c}& \colhead{d}& \colhead{e}
}
\startdata
$P$ [days]                    & 5.368483 (10$^{-5}$) &  12.91750 (0.002) & 66.845 (0.09) & 3.1485 (2 10$^{-4}$) \\
$e$                           & 10$^{-4}$(10$^{-3}$)$^{(*)}$ &   0.10    (0.05)$^{(*)}$  & 0.11   (0.08)$^{(*)}$ & 0.18 (0.08)$^{(*)}$   \\
$K$ [ms$^{-1}$]               & 12.49    (0.15)      &   3.40    (0.18)  & 1.90   (0.15) & 1.75   (0.14)   \\
$\omega$ [deg]                & 87.9     (10)        & 190.6     (7)     & 61.5   (8)    & 104.2  (7)      \\
$M_{0}$ [deg]                 & 190.3    (8)         & 193.3     (5)     & 353.9  (7)    & 144.7  (5)      \\
 \\
$m_{p} \sin i$ [M$_{\earth}$] & 15.61    (0.16)      & 5.71     (0.23)   & 5.4 (0.21)    & 1.805 (0.16)    \\
$a$ [AU]                      & 0.041                & 0.073             & 0.218         & 0.028           \\         
\hline
\hline
\\
New planets (V10)             & f	             & $g$            \\
\\
\hline
$P$ [days]                    & 440.7 (2.0)	     & 36.53 (0.32) \\
$e$                           & 0.17$^{(u)}$         & 0.19$^{(u)}$\\
$K$ [ms$^{-1}$]               & 1.18  (0.13)	     & 1.34  (0.17) \\
$\omega$ [deg]                & 330.8$^{(u)}$ 	     & 68.9$^{(u)}$\\
$M_{0}$ [deg]                 & 183.24 (10)	     & 248.1 (10)   \\
\\
$m_{p} \sin i$ [M$_{\earth}$] & 6.28 (0.65)	     & 3.18  (0.4) \\
$a$ [AU]                      & 0.767 	             & 0.146	   \\
\\
$\gamma_{\rm HARPS}$ [ms$^{-1}$] & -2.22 (0.2) \\
$\gamma_{\rm HIRES}$ [ms$^{-1}$] & 0.934 (0.2) \\
\hline
$\chi^2$                         & 520.5 \\
RMS [m/s]                        & 2.093 \\
\enddata
\label{tab:solution}
\tablenotetext{(u)}{Unconstrained parameter} 
\tablenotetext{(*)}{Compatible with 0}
\end{deluxetable}

\end{document}